\documentclass[aps,reprint,nofootinbib,groupedaddress,superscriptaddress]{revtex4-2}

\usepackage{amssymb, amsfonts, amsmath, bm}
\usepackage[
]{graphicx}
\usepackage{color}
\usepackage{mathrsfs}

\usepackage[
]{hyperref}
\hypersetup{%
 colorlinks=true,
  linkcolor=blue,   
  citecolor=blue,   
  urlcolor=blue     
}
\newcommand{\orcid}[1]{%
  \href{https://orcid.org/#1}{\includegraphics[height=0.7em]{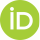}}%
}

\begin{document}
\title{Deflection Angle in the Strong Deflection Limit and Quasinormal Modes\\
in Stationary Axisymmetric Spacetimes}
\author{Takahisa Igata\:\!\orcid{0000-0002-3344-9045}
}
\email{takahisa.igata@gakushuin.ac.jp}
\affiliation{
Department of Physics, Gakushuin University,\\
Mejiro, Toshima, Tokyo 171-8588, Japan}
\date{\today}

\begin{abstract}
We derive a coordinate-invariant expression for the photon deflection angle in the strong deflection limit (SDL) of stationary axisymmetric spacetimes. The key logarithmic-divergence coefficient is shown to depend only on quantities locally measurable by a zero-angular-momentum observer---curvature scalars, the circumferential radius, and the proper angular velocity. The same coefficient governs the damping rate of quasinormal modes (QNMs) in the eikonal limit, establishing a curvature-based, model-independent connection between QNMs and lensing in the SDL near rotating compact objects. 
\end{abstract}

\maketitle

\textit{Introduction.}---%
Recent breakthroughs in black hole imaging~\cite{EventHorizonTelescope:2019dse,EventHorizonTelescope:2022wkp} and gravitational-wave astronomy~\cite{LIGOScientific:2016aoc,KAGRA:2021vkt} highlight the importance of clarifying the physics of unstable circular photon orbits (UCPOs) in strong-field regimes~\cite{Claudel:2000yi}. These orbits, a dynamical feature common to both observational windows, delineate the black hole shadow~\cite{Bardeen:1973tla,Luminet:1979}, govern the logarithmic divergence of the photon deflection angle in the strong deflection limit (SDL)~\cite{Bozza:2002zj,Tsukamoto:2016jzh}, and encode the frequency and damping of quasinormal-mode (QNM) ringdowns~\cite{Ferrari:1984zz,Mashhoon:1985cya,Kokkotas:1999bd,Berti:2009kk,Bolokhov:2025uxz}; see Refs.~\cite{Bolokhov:2025uxz} for a review. Quantitatively, each of these signatures depends only on two parameters---the critical impact parameter $b_{\mathrm{c}}$ and the second radial derivative of the null-geodesic effective potential $V''_{\mathrm{m}}$~\cite{Cardoso:2008bp,Stefanov:2010xz}. 

In static, spherically or axially symmetric spacetimes, a recent reformulation of the SDL has explicitly shown that the logarithmic-divergence coefficient $\bar{a}$ can be expressed solely in terms of $V''_{\mathrm{m}}$, and furthermore, that this quantity itself admits an expression constructed only from quantities measurable in a static frame---curvature scalars and the areal or circumferential radius~\cite{Igata:2025taz,Igata:2025plb}. These results may be regarded as a partial reformulation of the conventional, coordinate-dependent expressions for $\bar{a}$ in stationary axisymmetric spacetimes~\cite{Bozza:2002af}, thereby casting the coefficient into a fully coordinate-invariant form for this restricted static class. However, no equally local and coordinate-invariant formula is yet available for generic stationary spacetimes with frame dragging. Closing this gap is crucial not only for completing the SDL framework but also for clarifying the origin of eikonal-limit QNMs, whose damping is governed by the same underlying dynamical quantities~\cite{Stefanov:2010xz,Raffaelli:2014ola}.

To this end, we adopt the zero-angular-momentum-observer (ZAMO) frame together with the Newman--Penrose (NP) formalism for stationary, axisymmetric spacetimes. We obtain coordinate-invariant expressions for both the strong-deflection coefficient $\bar{a}$ and the eikonal-limit QNM damping rate: $\bar{a}$ depends solely on local curvature scalars, the circumferential radius, and the proper angular velocity, while the damping rate involves the same quantities in addition to a quasilocal redshift factor. This curvature-based formulation makes the connection between QNMs and lensing in the SDL fully coordinate- and model-independent, so that either observable can be used to infer---and cross-check---the local spacetime geometry.

Throughout this Letter, we use geometrized units with $G=1$ and $c=1$.

\textit{Geometry of spacetime.}---%
We consider a stationary and axisymmetric spacetime with two commuting Killing vector fields, $\partial_t$ and $\partial_\varphi$, which generate time translations and axial rotations, respectively. The metric $g_{ab}$ takes the canonical Weyl--Papapetrou form~\cite{Stephani:2003tm}
\begin{align}
\mathrm{d}s^2 =
&-\sigma\:\! e^{2\:\!\psi}\left(\mathrm{d}t+\sigma A\:\!\mathrm{d}\varphi\right)^2
\cr
&+ e^{-2\:\!\psi}\left[\:\!
e^{2\:\!\gamma}\left(\mathrm{d}\rho^2+\mathrm{d}\zeta^2\right)
+\sigma\:\!\rho^2\:\!\mathrm{d}\varphi^2\:\!\right],
\label{eq:metric}
\end{align}
where $\psi$, $\gamma$, and $A$ are functions of $\rho$ and $\zeta$. We set $\sigma=+1$ outside (and $-1$ inside) the static limit surface, defined by $e^{2\psi}\to 0$. We define the circumferential radius as the norm of the axial Killing vector $\partial_\varphi$; explicitly,
\begin{align}
R=e^{-\psi} \sqrt{\sigma\:\!(\rho^2-e^{4\psi}A^2)},
\end{align}
which remains regular across the static limit surface. We assume reflection symmetry about the equatorial plane $\zeta=0$, implying that each metric function is an even function of $\zeta$. Furthermore, we require the spacetime to be asymptotically locally flat on the equatorial plane, namely, $\psi, \gamma, A\to 0$ as $R\to \infty$.

We introduce ZAMOs, whose 4-velocity serves as the timelike basis vector of a tetrad~\cite{Bardeen:1972fi}:
\begin{align}
e_{(0)}^a=-\alpha^{-1}g^{ab}\nabla_b t
=\frac{1}{\alpha}\left(\partial_t^a+\Omega\:\! \partial_\varphi^a\right), 
\end{align}
where the redshift factor $\alpha \equiv 1/(-g^{tt})^{1/2}$ is given by
\begin{align}
\alpha =\frac{\rho}{R},
\end{align}
and the angular velocity with respect to the asymptotic rest frame is
\begin{align}
\Omega= \frac{e^{2\:\!\psi}A}{R^2}. 
\end{align}
The remaining three spacelike basis vectors, which together with $e_{(0)}^{a}$ complete the ZAMO tetrad, are
\begin{align}
e_{(1)}^a=e^{\psi-\gamma}\:\! \partial_\rho^a,
\quad
e_{(2)}^a=e^{\psi-\gamma}\:\! \partial_\zeta^a,
\quad
e_{(3)}^a=\frac{1}{R}\:\!\partial_\varphi^a.
\end{align}
The tetrad satisfies the orthonormality condition $g_{ab}e^{a}_{(\mu)}e^{b}_{(\nu)}=\eta_{(\mu)(\nu)}$, where $\eta_{(\mu)(\nu)} = \mathrm{diag}(-1,1,1,1)$. We raise tetrad indices with the inverse $\eta^{(\mu)(\nu)}$ and lower them with $\eta_{(\mu)(\nu)}$.

We focus on the Weyl tensor $C_{abcd}$---the trace-free part of the Riemann tensor---which is generally nonvanishing even for vacuum solutions of general relativity. Its tetrad components $C_{(\mu)(\nu)(\lambda)(\sigma)}$, evaluated in the ZAMO frame, can be decomposed into electric and magnetic parts as 
\begin{align}
E_{(i)(j)}&=C_{(i)(0)(j)(0)},
\\
B_{(i)(j)}&=\frac{1}{2}C_{(i)(0)}{}^{(k)(l)}\epsilon_{(k)(l)(j)(0)},
\end{align}
where $\epsilon_{(\mu)(\nu)(\lambda)(\sigma)}$ is the Levi-Civita symbol with $\epsilon_{(0)(1)(2)(3)} = 1$. 

We also consider the Einstein tensor $G_{ab}=R_{ab}-\tfrac{1}{2}\mathcal{R} g_{ab}$, which is constructed from the Ricci tensor $R_{ab}$ and the Ricci scalar $\mathcal{R}:=g^{ab}R^{ab}$. The tensor $G_{ab}$ encodes the matter distribution through the Einstein equations. Its tetrad components in the ZAMO frame are denoted by $G_{(\mu)(\nu)}$.

To facilitate the analysis of null directions and Weyl scalars, we introduce an NP null tetrad $\{l^a, n^a, m^a, \bar{m}^a\}$ constructed from the ZAMO tetrad. The real null vectors are defined by
\begin{align}
l^a=\frac{1}{\sqrt{2}}\big(e_{(0)}^a+ e_{(3)}^a\big),
\quad 
n^a=\frac{1}{\sqrt{2}}\big(e_{(0)}^a- e_{(3)}^a\big),
\end{align}
and the complex ones by
\begin{align}
m^a=\frac{1}{\sqrt{2}}\big(e_{(1)}^a+i \:\!e_{(2)}^a\big),
\quad
\bar{m}^a=\frac{1}{\sqrt{2}}\big(e_{(1)}^a-i \:\!e_{(2)}^a\big),
\end{align}
where $i$ is the imaginary unit. These vectors satisfy the standard normalization conditions $l^a n_a = -1$ and $m^a \bar{m}_a = 1$. All other contractions vanish. 

The complex Weyl scalars are defined by contracting the Weyl tensor with the null tetrad:\begin{align}
\Psi_0&=C_{abcd} l^a m^b l^c m^d,
\\
\Psi_1&=C_{abcd} l^a n^b l^c m^d,
\\
\Psi_2&=C_{abcd} l^a m^b \bar{m}^c n^d,
\\
\Psi_3&=C_{abcd} l^a n^b \bar{m}^c n^d,
\\
\Psi_4&=C_{abcd} n^a \bar{m}^b n^c \bar{m}^d. 
\end{align}
Among the NP Ricci scalars, we focus on
\begin{align}
\Phi_{00}&=\frac{1}{2} R_{ab} l^a l^b, 
\\
\Phi_{22}&=\frac{1}{2} R_{ab} n^a n^b.
\end{align}
According to the Einstein equations, these quantities are proportional to the energy densities measured by null observers with 4-momenta $l^a$ and $n^a$, respectively.

\textit{Unstable photon circular orbits.}---%
Due to reflection symmetry, geodesics initially confined to the equatorial plane remain in that plane. The Lagrangian governing photon motion in the plane $\zeta=0$ can be written as 
\begin{align}
\mathscr{L}
=\frac{1}{2}\Big[\!
-\sigma\:\!e^{2\:\!\psi}\!\left(\dot{t}+\sigma A\:\!\dot{\varphi}\right)^2\!
+ e^{-2\:\!\psi}\!\left(
e^{2\:\!\gamma}\dot{\rho}^2
+\sigma \rho^2\:\!\dot{\varphi}^2\right)
\Big]
\end{align}
with all functions evaluated at $\zeta=0$. An overdot denotes differentiation with respect to an affine parameter. In a stationary axisymmetric spacetime, the conserved energy $E$ and angular momentum $L$ along a photon orbit are given by 
\begin{align}
E&=e^{2\:\!\psi}\left(\sigma \:\!\dot{t}+A \:\!\dot{\varphi}\right),
\\
L&=\sigma\:\!e^{-2\:\!\psi} \rho^2 \dot{\varphi}-e^{2\:\!\psi}A \left(
\dot{t}+\sigma A\:\!\dot{\varphi}
\right).
\end{align}
Applying the null condition $\mathscr{L}=0$ and dividing by $\dot{\varphi}^2$ (assumed nonzero), we find
\begin{align}
\left(\frac{\mathrm{d}\rho}{\mathrm{d}\varphi}\right)^2+V=0,
\end{align}
where $V$ is the effective potential. Introducing the impact parameter $b\equiv L/E$, we obtain 
\begin{align}
V=\sigma e^{-2\:\!\gamma} \rho^2\left[\:\!
1-\frac{e^{-4\:\!\psi} \rho^2}{(b+\sigma A)^2}
\:\!\right].
\end{align}

Photon circular orbits are located at radii $\rho=\rho_{\mathrm{m}}^\pm$, which are determined by the simultaneous conditions $V=0$ and $V'=0$. These conditions yield the critical impact parameters
\begin{align}
b^{\pm}_{\mathrm{c}}=\sigma\left(\pm e^{-2\:\!\psi_{\mathrm{m}}^{\pm}} \rho_{\mathrm{m}}^{\pm}- A_{\mathrm{m}}^{\pm}\right),
\end{align}
with $\rho_{\mathrm{m}}^{\pm}$ satisfying
\begin{align}
\psi'=\frac{1\mp e^{2\:\!\psi} A'}{2\rho}.
\end{align}
Throughout, we adopt the convention that the upper (lower) signs correspond to prograde (retrograde) photon circular orbits. Hereafter, quantities with subscripts $\mathrm{m}$ and $\pm$ are evaluated at $\rho = \rho_{\mathrm{m}}^{\pm}$ on the equatorial plane.

The conditions for UCPOs at $\rho = \rho_{\mathrm{m}}^{\pm}$ are
\begin{align}
V''_{\mathrm{m},\pm}=&-2\sigma e^{-2\:\!\gamma_{\mathrm{m}}^{\pm}}\Big[\:\!
-1-2(\rho_{\mathrm{m}}^{\pm})^2 \psi''_{\mathrm{m},\pm}
\cr
&+e^{4\:\!\psi_{\mathrm{m}}^{\pm}}(A'_{\mathrm{m},\pm})^2
\mp e^{2\:\!\psi_{\mathrm{m}}^{\pm}} \rho_{\mathrm{m}}^{\pm} A''_{\mathrm{m},\pm}
\:\!\Big]<0.
\end{align}
We focus on the case $V''_{\mathrm{m},\pm}<0$ in what follows.

\textit{Deflection angle in the strong deflection limit.}---%
For a photon trajectory that reaches a turning point at $\rho=\rho_0$ and is scattered to infinity,
the total deflection angle is defined as
\begin{align}
\alpha(\rho_0) = I(\rho_0) - \pi,
\end{align}
with the orbit integral
\begin{align}
I(\rho_0)=2 \int_{\rho_0}^{\rho_{\infty}} \frac{|\mathrm{d}\rho|}{\sqrt{-V}}.
\end{align}
Here, $\rho_\infty$ denotes spatial infinity, where the circumferential radius $R$ diverges. The minimum radius $\rho_0$ is related to the impact parameter $b$ by the condition $V_0 = 0$, where $V_0$ is the effective potential evaluated at $\rho = \rho_0$. Hereafter, quantities labeled with the subscript $0$ are evaluated at $\rho = \rho_0$ on the equatorial plane. 

We particularly focus on the SDL, where the turning point $\rho_0$ approaches the UCPO radius $\rho_{\mathrm{m}}^{\pm}$. To isolate the resulting logarithmic divergence, we introduce the variable
\begin{align}
z=1-\frac{R_0}{R},
\end{align}
so that $z = 0$ corresponds to the closest approach. This choice of $z$ enables a systematic and coordinate-invariant treatment of the SDL, as first demonstrated in Ref.~\cite{Igata:2025taz}.

The leading divergence of $I(\rho_0)$ in the SDL is entirely captured by the integral 
\begin{align}
I_{\mathrm{D}}(\rho_0) = 2 \int_0^{1} \frac{\mathrm{d}z}{\sqrt{c_1 z+c_2 z^2}}, 
\end{align}
which isolates the contribution from the region near the UCPO. The coefficients $c_1$ and $c_2$ are given by
\begin{align}
c_1=-\frac{R'_0 V'_0}{R_0}, \quad c_2=-\frac{V''_0}{2}+3\left(
\frac{R'_0}{R_0}-\frac{R''_0}{2R'_0}
\right)V'_0.
\end{align}
The remainder, $I_{\mathrm{R}}(\rho_0)=I(\rho_0)-I_{\mathrm{D}}(\rho_0)$, remains finite. 

Since the turning-point condition $V_0 = 0$ links $b$ to $\rho_0$,
taking $\rho_0 \to \rho_{\mathrm m}^{\pm}$ is equivalent to
$b \to b_{\mathrm c}^{\pm}$. In the SDL, the deflection angle then exhibits a logarithmic divergence of the form
\begin{align}
\alpha \simeq -\bar{a}_\pm \log\left|
\frac{b}{b_{\mathrm{c}}^\pm}-1\right|,
\end{align}
where the coefficients $\bar{a}_\pm$ are given by
\begin{align}
\bar{a}_\pm=\sqrt{-\frac{2}{V''_{\mathrm{m},\pm}}}. 
\end{align}
The quantities $V''_{\mathrm{m},\pm}$ in $\bar{a}_\pm$ can be rewritten as 
\begin{widetext}
\begin{align}
V''_{\mathrm{m},\pm}
=\Big(\frac{R_{\mathrm{m}}^{\pm}}{1\pm v_{\mathrm{m}}^{\pm}}\Big)^2\left[\:\!
G_{(0)(0)}^{\mathrm{m},\pm}+G_{(3)(3)}^{\mathrm{m},\pm}-2(E_{(2)(2)}^{\mathrm{m},\pm}-E_{(1)(1)}^{\mathrm{m}, \pm})\pm (4B_{(1)(2)}^{\mathrm{m},\pm}+ 2G_{(0)(3)}^{\mathrm{m},\pm})
\:\!\right],
\label{eq:V''EG}
\end{align}
\end{widetext}
where 
\begin{align}
v_{\mathrm{m}}^{\pm}=\frac{R_{\mathrm{m}}^{\pm}\,\Omega_{\mathrm{m}}^{\pm}}{\alpha_{\mathrm{m}}^{\pm}}.
\end{align}
Note that $R_{\mathrm{m}}^{\pm}$ and the ratio $\Omega_{\mathrm{m}}^{\pm}/\alpha_{\mathrm{m}}^\pm$ can be measured locally by a ZAMO at $\rho=\rho_{\mathrm{m}}^\pm$. Hence, every term in Eq.~\eqref{eq:V''EG} is expressed entirely in locally measurable quantities.

Furthermore, using the NP scalars, the quantities $V''_{\mathrm{m},\pm}$ can be written for each branch as
\begin{align}
V''_{\mathrm{m},+}
=4\Big(\frac{R_\mathrm{m}^{+}}{1+ v_\mathrm{m}^{+}}\Big)^2 (\Psi_0^{\mathrm{m},+}+\Phi_{00}^{\mathrm{m},+}),
\\
V''_{\mathrm{m},-}
=4\Big(\frac{R_\mathrm{m}^{-}}{1- v_\mathrm{m}^{-}}\Big)^2 (\Psi_4^{\mathrm{m},-}+\Phi_{22}^{\mathrm{m},-}).
\end{align}
These two branches involve different NP scalars, reflecting the intrinsic chirality of the spacetime geometry as perceived by prograde and retrograde photons. In particular, $\Psi_0$ and $\Phi_{00}$ dominate the prograde branch, whereas $\Psi_4$ and $\Phi_{22}$ govern the retrograde one, highlighting the distinct curvature contributions to the instability of each orbit.

\textit{Quasinormal-mode frequencies.}---%
In the eikonal limit, QNMs are determined by the properties of the UCPOs. Specifically, the real part of the QNM frequency is the coordinate angular velocity $\Omega_{\mathrm{c}}^{\pm}$ of the prograde and retrograde branches, whereas the imaginary part (i.e., the damping rate) equals the decay rate $\lambda_{\mathrm{L}}^{\pm}$ of the corresponding unstable orbit. Thus, the QNM frequency can be written as
\begin{align}
\omega_{\mathrm{QNM}}^{\pm}=\Omega_{\mathrm{c}}^{\pm} l-i \left(n+\frac{1}{2}\right) \lambda_{\mathrm{L}}^{\pm},
\end{align}
where $l (\gg1)$ is the multipole index, and $n$ is the overtone number. The quantities $\Omega_{\mathrm{c}}^\pm$ are directly related to the critical impact parameters $b_{\mathrm{c}}^\pm$ as 
\begin{align}
\Omega_{\mathrm{c}}^\pm=\frac{1}{b_{\mathrm{c}}^\pm}=\frac{\alpha_{\mathrm{m}}^\pm(v_{\mathrm{m}}^\pm \pm1)}{R_{\mathrm{m}}^\pm}. 
\label{eq:Omegac}
\end{align}
The quantities $\lambda_{\mathrm{L}}^{\pm}$ are identified with the Lyapunov exponents~\cite{Cardoso:2008bp}. They take the form
\begin{align}
\lambda_{\mathrm{L}}^{\pm}=
|\Omega_{\mathrm{c}}^\pm|
\sqrt{-\frac{V''_{\mathrm{m},\pm}}{2}}.
\label{eq:Lyapu}
\end{align}
Thus, apart from the redshift factors $\alpha_{\mathrm{m}}^\pm$, which are quasilocal quantities, all quantities appearing in $\Omega_{\mathrm{c}}^\pm$ and $\lambda_{\mathrm{L}}^{\pm}$ are locally measurable by a ZAMO at $\rho=\rho_{\mathrm{m}}^\pm$.

From Eq.~\eqref{eq:Lyapu}, the relation between the SDL coefficient and the QNM parameters~\cite{Stefanov:2010xz,Raffaelli:2014ola} is given by
\begin{align}
\bar{a}_\pm=\frac{|\Omega_{\mathrm{c}}^\pm|}{\lambda_{\mathrm{L}}^{\pm}}. 
\end{align}
Finally, substituting Eq.~\eqref{eq:V''EG}, we obtain
\begin{widetext}
\begin{align}
\lambda_{\mathrm{L}}^{\pm}
=\frac{\alpha_{\mathrm{m}}^{\pm}}{\sqrt{2}}
\sqrt{\big|\:\!G_{(0)(0)}^{\mathrm{m},\pm}+G_{(3)(3)}^{\mathrm{m},\pm}-2(E_{(2)(2)}^{\mathrm{m},\pm}-E_{(1)(1)}^{\mathrm{m}, \pm})\pm (4B_{(1)(2)}^{\mathrm{m},\pm}+ 2G_{(0)(3)}^{\mathrm{m},\pm})\:\!\big|},
\end{align}
\end{widetext}
where we have used Eq.~\eqref{eq:Omegac}. The last term in the expression explicitly reflects the frame-dragging effect, which contributes differently to the prograde and retrograde modes through $B_{(1)(2)}^{\mathrm{m},\pm}$ and $G_{(0)(3)}^{\mathrm{m},\pm}$. Such contributions are absent in the static case and are a distinctive feature of rotation in stationary spacetimes.

Equivalently, the Lyapunov exponents take the following compact form in terms of the NP scalars:
\begin{align}
\lambda_{\mathrm{L}}^{+}= \alpha_{\mathrm{m}}^{+}\sqrt{2\:\!|\Psi_0^{\mathrm{m},+}+\Phi_{00}^{\mathrm{m},+}|},
\\
\lambda_{\mathrm{L}}^{-}=\alpha_{\mathrm{m}}^{-} \sqrt{2\:\!|\Psi_4^{\mathrm{m},-}+\Phi_{22}^{\mathrm{m},-}|}.
\end{align}
Note that $l^a_{\mathrm{m},+}$ ($n^a_{\mathrm{m},-}$) reduces to the tangent to the prograde (retrograde) UCPO. Therefore, $\Psi_0^{\mathrm{m},+}$ ($\Psi_4^{\mathrm{m},-}$) encodes the transverse local tidal field projected onto the screen space of $l^a_{\mathrm{m},+}$ ($n^a_{\mathrm{m},-}$), and $\Phi_{00}^{\mathrm{m},+}$ ($\Phi_{22}^{\mathrm{m},-}$) represents the null-energy density measured by the null observer $l^a_{\mathrm{m},+}$ ($n^a_{\mathrm{m},-}$) in general relativity. Note, however, that this identification is purely geometrical; the formulation does not rely on any specific field equations and therefore applies to any metric theory of gravity, including but not limited to general relativity.

In summary, the logarithmic-divergence coefficient in the SDL is expressed solely through coordinate-invariant quantities---curvature scalars, the circumferential radius, and the proper angular velocity---measurable by a ZAMO. Likewise, the damping rate of the QNMs in the eikonal limit is fixed by the same local inputs plus a quasilocal redshift factor. This curvature-based, coordinate-invariant bridge between lensing and ringdown observables provides a model-independent probe of strong-field gravity around rotating compact objects, ready for direct tests with VLBI images of black-hole shadows and next-generation gravitational-wave detectors. Full derivations will appear in a forthcoming paper; ongoing work extends the framework to QNM-relevant spherical photon orbits~\cite{Yang:2012he} and to their distinctive Weyl-curvature behavior in the extremal Kerr limit~\cite{Igata:2019pgb}, which are directly relevant to the slowly decaying (long-lived) quasinormal modes that emerge in the near-extremal Kerr spacetime~\cite{Konoplya:2017wot, Oshita:2022yry, Motohashi:2024fwt}.

~~

\begin{acknowledgments}
The author gratefully acknowledges the useful discussions with Yohsuke Takamori. This work was supported in part by JSPS KAKENHI Grants No.~JP22K03611, No.~JP23KK0048, and No.~JP24H00183 and by Gakushuin University. 
\end{acknowledgments}

~~

\textit{Data availability}---%
No data were created or analyzed in this article.

\end{document}